\documentclass[article]{elsarticle}

\usepackage{amssymb}
\usepackage{graphicx}
\usepackage{mathtools}
\usepackage{cryptocode}
\usepackage{underscore}
\usepackage{comment}
\usepackage{listings}

\begin{document}

\begin{frontmatter}



\title{Comment on ``Provably secure biometric-based client–server secure communication over unreliable networks"}


\author{Mahdi Nikooghadam}
\ead{mahdi.nikooghadam@aut.ac.ir}
\author{Hamid Reza Shahriari}
\ead{shahriari@aut.ac.ir}
\cortext[mycorrespondingauthor]{Hamid Reza Shahriari}
\address{Department of Computer Engineering, Amirkabir University of Technology, Tehran, Iran}

\begin{abstract}
In key agreement protocols, the user will send a request to the server and the server will respond to that message. After two-way authentication, a secure session key will be created between them. They use the session key to create a secure channel for communication. In 2021, Saleem et al. proposed a protocol for securing user and server communications, claiming that their proposed protocol meets a variety of security needs and is also resistant to known types of attacks. In this article, we will show that Saleem et al.'s scheme does not meet the security requirement of perfect forward secrecy.
\end{abstract}




\end{frontmatter}


\section{Introduction}
\label{intro}
Currently, with the significant growth of information technology, one of the major concerns, especially in the field of IoT, is about authentication protocols and the security of messages that send between users and server.\\
There is a lot of research present to ensure the information security requirements of messages that transmit between users and servers. Most research is based on smart cards.\\
In 1981, Lamport et al. \cite{lamport} proposed a simple authentication protocol based on one-way hash functions.\\
However, Sung-Ming et al. \cite{Sung-Ming} proved that the protocol presented in \cite{lamport} is not resistant to stolen verifier attack. He showed the attacker can obtain users' confidential information by infiltrating the server database. Also in 2000, a key authentication scheme was proposed by  Hwang et al. \cite{Hwang}. The protocol has a very high computational and communication overhead.\\
Das presented a two-factor authentication protocol for communication between users and servers in IoT environment based on a password and smart card\cite{Das}. The study by Khan, Alghathbar\cite{Khan} and  Lee, et al. \cite{Lee} demonstrates that the Das protocol does not guarantee user anonymity and is susceptible to password guessing attacks.In 2016, Nikoghadam and Arshad \cite{arshad} proposed a protocol for authentication and key agreement with the aim of ensuring user anonymity. This protocol could not meet the perfect forward secrecy\cite{perfectmorteza}.\\

An improved two-factor authentication technique for client-server environments was presented by Xie et al. \cite{Xie}  in 2014 . After then, Lu et al. \cite{Lu} conducted cryptanalysis on \cite{Xie} and discovered that it was susceptible to tracking, insider attack, and user impersonation attacks. In order to fix the problems with the Xie et al. protocol, Lu et al. propose an improved two-factor authentication scheme. The protocol \cite{Lu} fails to withstand user impersonation, server impersonation, and MITM attacks, according to Mahmood et al. in 2020 \cite{Mahmood}. In order to address the shortcomings of \cite{Lu}, Mahmood et al. introduced an anonymous authentication mechanism for client-server environments. However, Mahmood et al scheme were not suitable for real-world use.

Saleem et al\cite{salman2021} also presented a scheme in 2021 for authentication and key agreement to secure the communications between servers and users. They claimed the proposed protocol could meet various security requirements and was also resistant to known attacks. In this paper, we show that Saleem et al.'s scheme does not meet perfect forward secrecy.

\section{Overview and cryptanalysis }
\label{sec:salman}
In this section, we carry out cryptanalysis of Saleem et al.’s scheme and show it is vulnerable against perfect forward secrecyy~\cite{perfectsat}~\cite{perfectmorteza}~\cite{perfectmahdi}, Also, we show the Saleem et al.’s scheme can’t provide security for authentication and key agreement protocols. 
\subsection{Overview of Saleem et al.'s Scheme}
\label{sec:OAnalysis}
Salman et al. scheme includes two main phases, authentication and key agreement. In registration phase, the communication channel between active entities considers secure. \\ The communication channel is also regarded as insecure in the authentication and key agreement phase, so the attacker can eavesdrop, manipulate, or store messages during this phase. Table 1 shows the notations used in Salman et al.’s scheme. Also, figures 1 and 2 show their proposed protocol in detail.
\begin{table}
\caption{Notations used in Saleem et al.'s scheme~\cite{salman2021}}
\label{tab:KN}       
\begin{tabular}{ll}
\hline\noalign{\smallskip}
Notation & Description \\
\hline\noalign{\smallskip}
\emph{$C_c$} & $c$th client \\
\emph{$S_s$} & $s$th server \\
\emph{$ID_c$} & Identity of $C_c$ \\
\emph{$PW_c$} & Password of $C_c$ \\
\emph{$B_c$} & Biometric of $C_c$ \\
\emph{$E_p(a,b)$} & Elliptic curve over $Z_p$ \\
\emph{$P$} & Base point of  $E_p(a,b)$ \\
\emph{$s$} & Secret key of $S_s$, where $s \in Z_p^{*}$ \\
\emph{$Pub$} & Public key of $S_s$, where $Pub=s.P$ \\
\emph{$h(.)$} & Secure hash function \\
\emph{$SC_c$} & Smartcard  of $C_c$ \\
\emph{$A$} & A malicious client/adversary \\
$\oplus ,||$ & XOR/Concatenation operator \\
$SK$ & Common session key between \emph{$C_c$} and \emph{$S_s$} \\
\noalign{\smallskip}\hline\noalign{\smallskip}
\end{tabular}
\end{table}

\begin{figure}
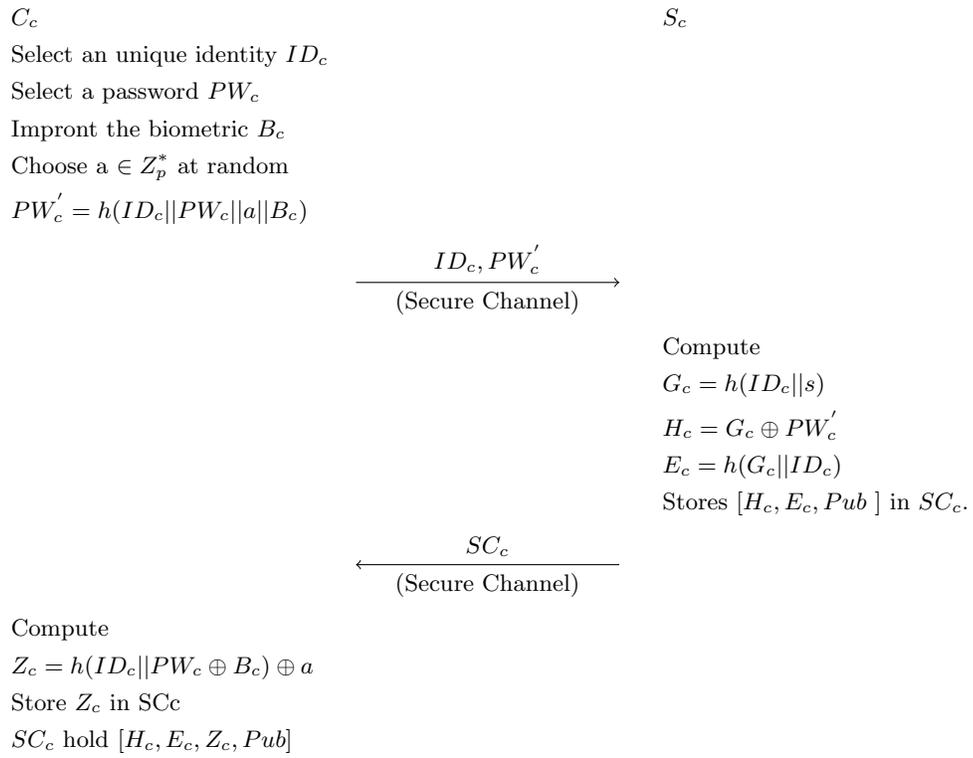

\procedure{Registration Phase}{%
\textbf{$ C_c$} \> \> \textbf{$S_c$}  \\
\text{Select  an unique identity $ID_c$ }\\
\text{Select a password $PW_c$}\\
\text{Impront the biometric $B_c$}\\
\text{Choose a $\in Z_p^{*} $ at random}\\
\text{$PW_c^{'}=h(ID_c||PW_c||a||B_c)$}\\
 \> \sendmessageright{top=\text{$ID_c,PW_c^{'}$} , bottom=\text{(Secure Channel)}} \> \\
\> \>\text{Compute}\\
\> \> \text{$G_c=h(ID_c||s)$}\\
\> \> \text{$H_c=G_c\oplus PW_c^{'}$} \\
\> \> \text{$E_c=h(G_c||ID_c)$} \\
\> \> \text{Stores [$H_c,E_c,Pub$ ] in $SC_c$.} \\
\> \sendmessageleft{top=\text{$SC_c$} , bottom=\text{(Secure Channel)}} \> \\
\text{Compute}\\
\text {$Z_c=h(ID_c||PW_c\oplus B_c)\oplus a$}\\
\text {Store $Z_c$ in SCc}\\
\text {$SC_c$  hold [$H_c,E_c,Z_c,Pub$]}
\\
}
\caption{Registration Phase of Saleem et al.’s scheme }
\end{figure}
\begin{figure}
\procedure{Login and Authentication Phase}{
\textbf{$C_c$} \> \> \textbf{$S_s$}  \\
\text {Enter $ID_c$ and $PW_c$}\\
\text{Imprint the biometric $B_c$}\\
\text{$a=Z_c \oplus h(ID_c||PW_c \oplus B_c )$}\\
\text{$G_c^{'}=H_c \oplus h(ID_c||PW_c||a||B_c)$} \\
\text{$E_c^{'}=h(G_c{'}||ID_c)$}\\
\text{If $(E_c^{'}\ne E_c)$} \\
\text{Abort  the session} \\
\text{Else, choose $r_c \in Z_p^{*} at random$} \\
\text{Compute} \\
\text{$M_c=r_c.Pub=r_c.s.P$} \\
\text{$PID_c=ID_c \oplus r_c.P$} \\
\text{$N_c=r_c \oplus h(E_c||t_c)$ } \\
\text{$Auth_c=h(ID_c||G_c||r_c||t_c)$ } 
\> \sendmessageright{top=\text{${M_c,PID_c,Auth_c,N_c,t_c}$}} \> \\
\> \>
\text{if  $(t_s-t_c>\Delta t)$}\\
\> \> \text{Abort the session} \\
\> \> \text{Else and compute } \\
\> \> \text{$ID_c=PID_c \oplus s^{-1}.M_c$}\\
\> \> \text {$G_c=h(ID_c||s)$}\\
\> \> \text {$E_c=h(G_c||ID_c)$}\\
\> \> \text {$r_c=N_c \oplus h(E_c||t_c)$}\\
\> \> \text {$Auth_c^{'}=h(ID_c||G_c||r_c||t_c)$}\\
\> \> \text {if $Auth_c^{'} \ne Auth_c$}\\
\> \> \text {Terminate the session}\\
\> \> \text{else,choose $r_s \in Z_p^{*}$ at random} \\
\> \> \text{compute $O_s=r_s \oplus r_c$} \\
\> \> \text{$SK=h(G_c||r_c||r_s||t_c||t_s)$} \\
\> \> \text{$Auth_s=h(SK||E_c||ID_c)$}\\
\> \sendmessageleft{top=\text{$O_s,Auth_s,t_s$} , bottom=\text{}} \> \\
\text{if  $(t_k-t_s>\Delta t)$} \\
\text{Abort the session} \\
\text{Else and compute $r_s=O_s \oplus r_c$}\\
\text{Computes $SK=h(G_c^{'}||r_c||r_s||t_c||t_s)$} \\
\text {if ($Auth_s^{'} \ne h(SK||E_c||ID_c)$)}\\
\text{Abort the session} \\
\text{Else , accept SK is the session key} \\
}
\caption{Authentication phase of Saleem et al.’s scheme }
\label{fig:PAut}
\end{figure}

\subsection{Cryptanalysis of Saleem et al.'s Scheme}
In this section, we analyze the defects of the proposed scheme by Saleem et al. And explain its vulnerabilities in detail.
\label{sec:KAnalysis}
In this section, we analyze Saleem et al.'s scheme and explain its vulnerabilities in detail.  Saleem et al. claimed that their protocol could meet this forward secrecy. We prove the claim of Saleem et al is incorrect and their protocol can not guarantee these security requirements. \\
 In perfect forward secrecy, it is assumed that if an attacker has access to long-term parameters of active entities such as private and public keys, he should not be able to access the session key.\\

Assuming this, if the attacker obtain access to the server’s private key, a premise of perfect forward secrecy, since the  $M_c$ and $PID_c$ parameters are exchanged on the public channel , the attacker can obtain the $ID_c$ parameter through the  $ID_c = PID_c\oplus s^{-1}.M_c$.\\ 

In the next step, by the assumption that the attacker has server private key, and also  $ID_c$ which was obtained in the previous step, it will be able to obtain $G_c$ from the relation $G_c=h(ID_c||s)$ Who has already obtained $ID_c$ will be able to generate $E_c$ from $E_c=h(G_c||ID_c)$.\\

Now considering that the parameter $N_c$ and the time stamp  $t_c$ are exchanged on the public channel, as well as the parameters that the attacker has obtained so far, he/she will be able to generate $r_c$ through $r_c=N_c\oplus h(E_c||t_c)$.\\
Next, by considering the parameters that exchange in the second message in the authentication and key agreement phase $(O_s, Auth_s, t_s)$ and the assumption that the attacker can distinguish them, he will be able to obtain the $r_s$ parameter from $O_s=r_s\oplus r_c$\\
Finally, considering all the parameters obtained so far, it is possible to generate the session key from the $SK=h(G_c||r_c||r_s||t_c||t_s)$. In this way,  we show the claim of Saleem et al. \cite{salman2021} is incorrect.

\section{Conclusion and Future Work}
\label{sec:Conclusion}
Nowadays providing a secure communication channel that enables users to send messages to servers with preservation of privacy, has been considered by many researchers.In this article, we review the Saleem et al scheme, and we showed that even though they use elliptic curve encryption in their proposed protocol, their scheme does not provide perfect forward secrecy.\\
Their proposed protocol can't meet perfect forward secrecy because elliptic curve encryption was not used properly. In future work, we will propose a secure key authentication scheme using elliptic curve encryption and the ECDHP. The protocol will meet the security requirements in authentication and key agreement protocols.





\end{document}